\documentclass[aps,prl,preprint,showpacs]{revtex4}

\usepackage{graphicx}
\usepackage{dcolumn}
\usepackage{bm}

\begin{document}


\title{Interplay of electronic correlations and lattice instabilities in BaVS$_3$}

\author{Kwang-Yong Choi,$^{1,2}$ Dirk Wulferding,$^{2}$ Helmuth Berger,$^3$ and Peter Lemmens$^2$}
\affiliation{Department of Physics, Chung-Ang University, 221 Huksuk-Dong, Dongjak-Gu,
Seoul 156-756, Republic of Korea}
\affiliation{Institute for Condensed Matter Physics, TU Braunschweig, D-38106 Braunschweig, Germany}
\affiliation{Institute de Physique de la Mati$\acute{e}$re Complexe, EPFL, CH-1015 Lausanne, Switzerland}

\date{\today}
\begin{abstract}

The quasi-one-dimensional metallic system BaVS$_3$ with a metal-insulator transition at $T_{MI}=70$~K shows large changes of the optical phonon spectrum, a central peak, and an electronic Raman scattering continuum that evolve in a three-step process. Motivated by the observation of a strongly fluctuating precursor state at high temperatures and orbital ordering and a charge gap at low temperatures we suggest a concerted action of the orbital, electronic, and lattice subsystems dominated by electronic correlations.

\end{abstract}


\maketitle



The discussion on the relation of lattice distortions and electronic correlations effects has recently been refueled by the good agreement that certain correlated electron models show in calculating structural relaxations of transition metal compounds~\cite{leonov}. This implies in some sense a superiority of electronic with respect to lattice degrees of freedom. On the other side, the considered system, KCuF$_3$, is a high symmetry compound with a single hole in an e$_g$ state. This realizes a comparatively simple and clear electronic and structural situation. The lattice distortion, as a cooperative Jahn-Teller transition, happens at very high temperatures and the related orbital orbital ordering leads to its ambient temperature, low energy description~\cite{Fulde} as a one dimensional spin s=1/2 chain system. However, correlated electron systems with metal-insulator transitions (MIT) accompanied by Mott localization and charge ordering~\cite{Imada} are wellknown for more complex electronic systems involving multiple orbitals. They notoriously tend to complex structural distortions and a low symmetry crystallographic ground states. It is therefore of capital importance to perform further studies on the relation of structural and electronic degrees of freedom on such a more complex correlated electron systems.

In this context, the quasi-one-dimensional (quasi-1D) metallic system BaVS$_3$ is a fascinating example to study cooperative spin, charge, and orbital correlations with reduced dimensionality, highlighted by the observation of a Luttinger-liquid behavior~\cite{Imada} and focussed theoretical studies~\cite{Penc02,Lechermann07}. In spite of extensive experimental and theoretical efforts on this system, however, there is no consensus on the phase diagram and the hierarchies of the involved energy scales~\cite{Fazekas}.

In our Raman scattering study, we provide direct evidence for an orbital-ordered and charge-gapped ground state of BaVS$_3$. In addition, the existence of a strongly fluctuating state in the metallic phase and a unstable state in the insulating phase suggests an intimate coupling of the different subsystems.

%

BaVS$_3$ crystallizes in the hexagonal structure (P6$_3$/mmc with Z=2) at room temperature~\cite{Gardner}.
The VS$_3$ spin chains are formed by stacking face sharing VS$_6$ octahedra along the \emph{c} axis.
Each chain is separated by Ba atoms in the {\it ab} planes, realizing a quasi-1D structure.
Two electrons in the 3d $t_{2g}$ atomic levels are split into a broad A$_{1g}$ band ($d_{z^2}$ orbital) and two quasidegenerate narrow e($t_{2g}$) bands in the trigonal crystal field. The former is responsible for the 1D metallic behavior while the latter dictate localized 3D electrons~\cite{Lechermann07,Whangbo,Lechermann05,Lechermann06}.
Recent angle resolved photoemission spectroscopy (ARPES)~\cite{Mitrovic07} and local density approximation (LDA) combined with dynamical mean-field theory~\cite{Lechermann07} have highlighted the significance of large on-site Coulomb repulsion and Hund's rule coupling leading to nearly half-filling of both A$_{1g}$ and E$_{g1}$ bands. With decreasing temperature BaVS$_3$ undergoes three successive phase transitions: First, a second-order structural phase transition to the orthorhombic $Cmc2_1$ space group at $T_S$=240~K with distorted V$^{4+}$ zigzag chains. Second, a MIT at $T_{MI}$=70~K accompanied by a symmetry lowering to a monoclinic phase ($Im$), a sharp maximum of the magnetic susceptibility and specific heat anomalies~\cite{Imai96}. Third, magnetic ordering for T$<$$T_X=30$~K.

The main driving force of the MIT is proposed to be a Peierls-like transition stabilizing a $2k_F(A_{1g})$ CDW with a critical wave vector $q_c=c^{\ast}/2$, precipitated by huge precursor fluctuations at about 170~K~\cite{Fagot03}.  Remarkably, an internal distortion of the VS$_6$ octahedra and a tetramerization of the V$^{4+}$ chains occur without appreciable charge disproportionation~\cite{Fagot06}. This was discussed in terms of an orbital ordering in the insulating phase. The orbital-ordering scenario was supported by several experimental investigations~\cite{Fagot06,Nakamura97,Mihaly00,Ivek} and theoretical modelling~\cite{Lechermann07} although direct evidence is still lacking.

Raman spectroscopy is a tool of choice to address these issues because optical phonons and electronic scattering can serve as a local probe to explore the evolution of the orbital, electronic and lattice subsystem.
The used single crystals of BaVS$_3$ grown by the Tellurium flux method~\cite{Mihaly00} had typical dimensions of $3\times 0.2\times0.2~\mbox{mm}^3$. Raman scattering experiments were performed in quasi-backscattering using the $\lambda= 488$~nm excitation line with a power of 5~mW focused on 0.1~mm diameter.


\begin{figure}[tbp]
\linespread{1}
\par
\includegraphics[width=7cm]{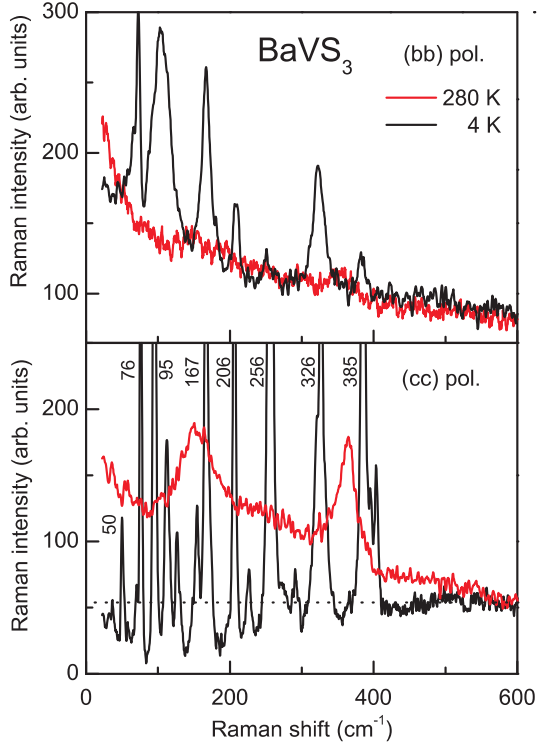}
\par
\caption{Raman spectra of BaVS$_{3}$ for light polarizations
parallel (cc) and perpendicular (bb) to the chain direction at 4~K and room temperature,
respectively. The numbers denote the frequency of principal phonon peaks at 4~K.} \label{fig:1}
\end{figure}

Figure~1 compares Raman spectra for polarization parallel (cc) and perpendicular (bb) to the chain direction at 4~K and 280~K, respectively. In the high-temperature metallic phase two peaks around 157 and $365~
\mbox{cm}^{-1}$ are superimposed by an electronic continuum. They belong to four symmetry-allowed modes; $\Gamma_{Raman}=1A_g(xx,yy,zz)+3E_{2g}(xx-yy,xy)$
~\cite{Popovic}.  The $157~
\mbox{cm}^{-1}$ E$_{2g}$ and $365~\mbox{cm}^{-1}$ A$_{1g}$  modes correspond to a vibration of the S atoms and a breathing mode of the VS$_6$ octahedra, respectively. These modes are weak and broad due to the screening by conduction electrons. In the low-temperature insulating phase we observe 8 rather broad A$_1$ modes with inter-chain polarization.
This agrees well with the factor group prediction $\Gamma_{Raman}=8A_1(xx,yy,zz)$ for the space group $Cmc2_1$. In in-chain polarization weak satellite peaks are observed in addition to the sharp, intense principal modes. The electronic background shows a drastic change as well. The markedly different polarization dependence corresponds to anisotropic electronic polarizabilities in the quasi-1D nature of BaVS$_3$.

\begin{figure}[tbp]
\linespread{1}
\par
\includegraphics[width=6.5cm]{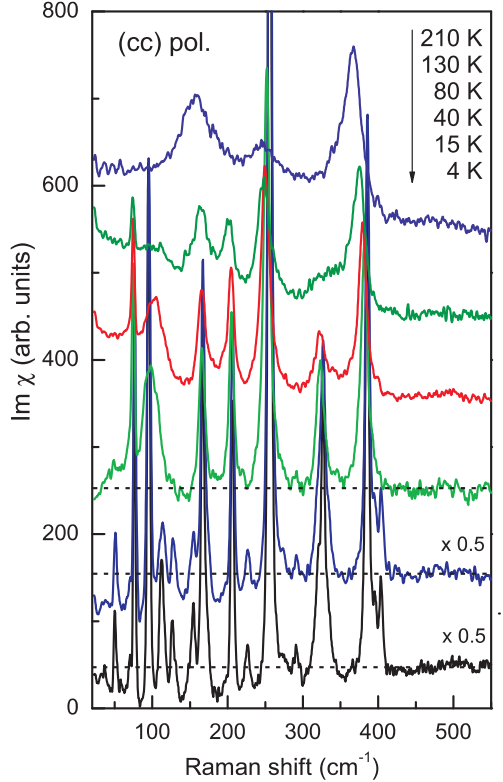}
\par
\caption{Temperature dependence of the Raman response $\mbox{Im}\chi$ with
in-chain polarization. The dashed lines are the scattering background.
The data are systematically shifted and multiplied by the given factor.} \label{fig:2}
\end{figure}

Shown in Fig.~2 is the detailed temperature dependence of the Raman response $\mbox{Im}\chi$,
which is corrected by the Bose thermal factor $[1+n(\omega)]
=[1-\exp(-\hbar\omega/k_BT)]^{-1}$ from the measured Raman scattering intensity.
With decreasing temperature the electronic scattering is suppressed and shows a depletion of
spectral weights below $\sim 400$ cm$^{-1} (=49.6~\mbox{meV})$ for $T<40$~K, indicative of
the opening of a charge gap.  The gap obtained from Raman measurements falls in the same range as in optical conductivity ($\Delta_{ch}\sim 42$~meV)~\cite{Kezsmarki}, photoemission ($\Delta_{ch}=60 -70$~meV)~\cite{Mitrovic07}, and resistivity measurements ($\Delta_{ch}\sim 52$~meV)~\cite{Mihaly00}. We note that in the ARPES and optical conductivity experiments a pseudogap feature is visible up to $\sim 80$~K. In contrast, the charge gap is discernible in our Raman spectra for $T <T_X$. This is because the pseudogap feature is covered by other low-energy excitations above $T_X$ (see below).

\begin{figure}[tbp]
\linespread{1}
\par
\includegraphics[width=6.5cm]{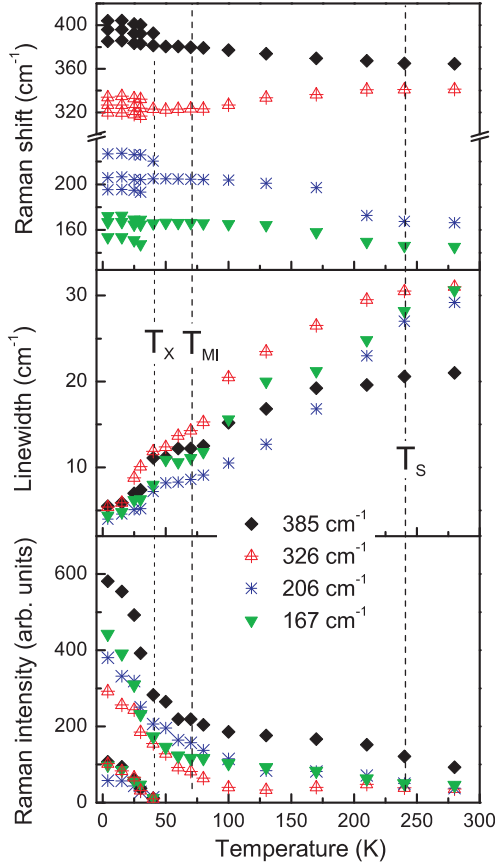}
\par
\caption{Temperature dependence of the peak frequencies (upper panel),
the linewidth (middle panel), and intensity (lower panel) of the 167, 206,
326, and 385 cm$^{-1}$ modes. The vertical dashed lines denote the consecutive
phase transitions.
} \label{fig:3}
\end{figure}

The phonon modes exhibit a number of interesting and anomalous changes in frequency and linewidth as a function of temperature.  For a quantitative analysis we fit them to Lorentzian profiles. The resulting frequency, linewidth, and intensity of the 167, 206, 326, and 385 cm$^{-1}$ modes are summarized in Fig.~3. The other modes have essentially the same behavior (not shown here).

Upon cooling from $T_S$ to $T_{MI}$ the phonon energies undergo a giant energy shift by $10-40~ \mbox{cm}^{-1}$. Also their linewidths decrease enormously whereas their intensities show no substantial change. Between $T_{MI}$ and $T_X$ the phonon modes exhibit sizable changes in both intensity and linewidth. In contrast, the phonon energies hardly vary with temperature. Below $T_X$ the intensities increase abruptly and the linewidths drop rapidly. The most salient feature is a splitting of the phonon modes. Two new peaks show up as a shoulder of each principal mode with weaker intensities. In addition, all 8~A$_1$ modes have the same three-peak feature. This suggests that the new modes are due to a zone folding of zone boundary modes to the $\Gamma$ point. The two zone-folded modes imply the presence of a new periodicity with four atomic spacings known as a tetramerization of V$^{4+}$ chains~\cite{Fagot06}. Since the zone-folded modes are clearly resolved only with in-chain polarization they should be ascribed to the $4k_F$ effects of the essentially 1D metallic bands. However, a conventional CDW does not give rise to a backfolding of all phonon branches. Rather, our results are compatible with a periodic orbital arrangement of the four V$^{4+}$ atoms, which induces strong variations of electronic polarizabilities. Therefore, we give spectroscopic evidence for a static orbital ordering.

\begin{figure}[tbp]
\linespread{1}
\par
\includegraphics[width=8cm]{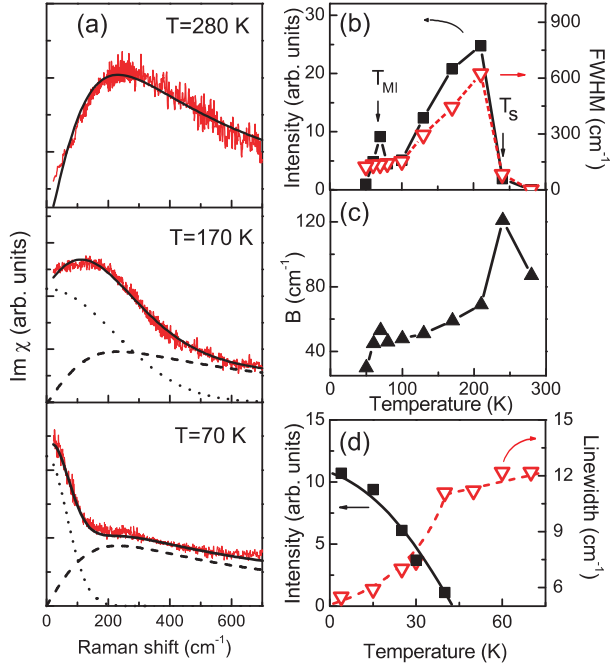}
\par
\caption{ (a) Electronic Raman response at 280, 170, and 70~K obtained after subtracting phonon peaks.
The fitting corresponds to a collision-dominated model (dashed lines), a Gaussian profile (dotted) and the sum of the two contributions (solid). (b)  Temperature dependence of the scattering amplitude (full squares) and full-width at half maximum (inverse open triangle) extracted from the Gaussian fit.
(c) Temperature dependence of the scattering amplitude obtained by a collision-dominated model.
(d) Temperature dependence of the intensity of the 385 cm$^{-1}$
zone-folded mode (full square) and the linewidth of  the 385 cm$^{-1}$
principal mode (open inverse triangle). Solid and dashed lines are guides to the eye.
} \label{fig:4}
\end{figure}

Next we will focus on the evolution of the background scattering. In the high temperature metallic phase the electronic Raman response is well fitted by a collision-dominated (CD) model, $\mbox{Im}\chi
\propto B\omega\Gamma/(\omega^2+\Gamma^2)$, where B is the scattering amplitude and $\Gamma$ is the carrier scattering rate. Approaching $T_S$ a broad elastic-scattering maximum becomes prominent, which is described by a Gaussian profile. This is a central peak arising from the second-order structural phase transition~\cite{Choi}.  We find that in a wide temperature range of $50 < T <T_S$ the background response is well fitted by the sum of the CD scattering (electronic response) and the Gaussian profile (cental peak). Representative fits are displayed in Fig.~4(a) with fitting parameters summarized in Figs.~4(b) and (c). The intensity of the central peak becomes stronger upon cooling through $T_S$, shows a maximum around 210~K and then decreases rapidly with a small maximum at $T_{MI}$, consistent with the second structural phase transition.  Both its linewidth and the amplitude of the electronic response resemble the temperature dependence of the intensity. Surprisingly, no well-defined soft mode shows up below $T_S$. Instead, the central peak persists well below $T_S$ with an exceptionally broad full-width at half maximum (FWHM) of several hundreds $\mbox{cm}^{-1}$. This feature indicates the presence of electronic and structural fluctuations in nearly the whole temperature range.

In the following we will reflect on the underlying mechanism of the MIT. The high-temperature metallic state ($T>T_S$) is distinguished from simple metals due to the CD scattering with a carrier scattering rate $\Gamma=230~\mbox{cm}^{-1}$. The metallic A$_{1g}$ electrons alone cannot yield such a low-energy electronic Raman response due to screening by itinerant electrons~\cite{Zawadowski}. Therefore, the presence of the CD response means that there exist strong electronic scattering from spin(orbital) fluctuations of the localized E$_g$ electrons. This signals that electronic correlations are a driving factor of the manifold MIT.


In the intermediate-temperature metallic phase ($T_{MI}<T<T_S$) there appears a strong renormalization of the phonon energies, a three-times decrease of the phonon linewidths, and a suppression of the central peak and CD response with decreasing temperature. In our case, the phonon linewidth is determined by an electron-phonon mechanism and thus is a measure of the free carrier concentration. Our observation indicates a continuous reduction of the electronic density of states at the Fermi level $E_F$, which diminishes the central peak and CD scattering amplitudes [compare the middle panel of Fig.~3 with Figs.~4(b) and (c)]. Our results unveil that an electronic instability starts around $T_S$ and develops gradually. This is fully consistent with the persistence of the fluctuating $2k_F(A_{1g})$ CDW superstructure up to 170~K~\cite{Fagot03}. The evolution of the central peak provides complimentary information about the lattice instability because it is associated with the decay of a soft mode into acoustic phonons or phonon density fluctuations. The enduring large linewidth down to 50~K indicates persisting structural fluctuations. Together with the similar temperature dependence of the CD response due to localized E$_g$ electrons we conclude that the Fermi surface, structural (orbital), and lattice instabilities develop consonantly and their strong fluctuations characterize the precursor transition regime.

In the low-temperature insulating phase ($T_{MI}<T<T_X$) there is no shift of the phonon energies. In contrast, the phonon linewidths decrease moderately and the central peak and CD scattering amplitudes drops rapidly just after reaching a maximum at $T_{MI}$. The zone-folded modes related to a $2k_F(A_{1g})$ CDW instability is not seen. This is consistent with the lack of a magnetic long-range order between $T_{MI}$ and $T_S$~\cite{Mihaly00}. Through the MIT the orbital, electronic, and structural states become quasi-static but not totally frozen-in. In this regime, several metastable states compete with each other.

For $T<T_{X}$, there appear the zone-folded modes associated with the $4k_F$ instability, which is a measure of the orbital order parameter. Their intensity increases strongly upon cooling from 40~K [see Fig.~4(d)]. At the same time, the cental peak and CD electronic scattering disappear. Instead, the deletion of the low-frequency electronic scattering evidences the opening of a charge gap. This is further supported by an exponential-like drop of the linewidth, which means a depletion of the density of states at $E_F$.  The fact that the onset temperature of the superstructure peak and the drop of the linewidth coincide  corroborates that the  orbitals are important in stabilizing the charge-gapped insulating ground state.

To conclude, we demonstrate that in BaVS$_{3}$ electronic fluctuation exist already above the phase transitions. We find that the exceptionally rich phases are initiated by electronic correlations. These effects in turn cause the entangled instability of the orbital, electronic, and lattice subsystem. Thus, the "order parameter" of the MIT in this compound is a complex quantity involving all three subsystems.

This work was supported by the DFG, the ESF program \emph{Highly Frustrated Magnetism}, the Swiss NSF and the NCCR MaNEP. KYC acknowledges financial support from the Alexander-von-Humboldt Foundation.

\end{document}